\input harvmac
\noblackbox

\font\cmss=cmss10 \font\cmsss=cmss10 at 7pt
 \def\inbar{\,\vrule height1.5ex width.4pt depth0pt}
\def\IZ{\relax\ifmmode\mathchoice
{\hbox{\cmss Z\kern-.4em Z}}{\hbox{\cmss Z\kern-.4em Z}}
{\lower.9pt\hbox{\cmsss Z\kern-.4em Z}}
{\lower1.2pt\hbox{\cmsss Z\kern-.4em Z}}\else{\cmss Z\kern-.4em
Z}\fi}
\def\IB{\relax{\rm I\kern-.18em B}}
\def\IC{{\relax\hbox{$\inbar\kern-.3em{\rm C}$}}}
\def\ID{\relax{\rm I\kern-.18em D}}
\def\IE{\relax{\rm I\kern-.18em E}}
\def\IF{\relax{\rm I\kern-.18em F}}
\def\IG{\relax\hbox{$\inbar\kern-.3em{\rm G}$}}
\def\IGa{\relax\hbox{${\rm I}\kern-.18em\Gamma$}}
\def\IH{\relax{\rm I\kern-.18em H}}
\def\II{\relax{\rm I\kern-.18em I}}
\def\IK{\relax{\rm I\kern-.18em K}}
\def\IC{\relax{\rm I\kern-.18em C}}
\def\IR{\relax{\rm I\kern-.18em R}}


\lref\garhir{G. Horowitz and H. Ooguri, ``Spectrum of Large N Gauge
Theory from Supergravity,'' hep-th/9802116.}
\lref\malda{J. Maldacena, ``The Large N limit of Superconformal Field
Theories and Supergravity,''
hep-th/9711200.}
\lref\gubkleb{S. Gubser and I. Klebanov, ``Absorption by Branes and
Schwinger Terms in the World Volume Theory,'' Phys. Lett. {\bf B413}
(1997) 41.}
\lref\gubklebp{S. Gubser, I. Klebanov and A. Polyakov, ``Gauge Theory
Correlators from Non-critical String Theory,'' hep-th/9802109.}
\lref\stanford{P. Claus, R. Kallosh, and A. van Proeyen, ``M Fivebrane
and Superconformal (0,2) Tensor Multiplet in Six 
Dimensions," hep-th/9711161\semi
R. Kallosh, J. Kumar and A. Rajaraman, ``Special Conformal Symmetry of
Worldvolume Actions,'' hep-th/9712073\semi
P. Claus, R. Kallosh, J. Kumar, P. Townsend and A. van Proeyen, ``Conformal
Theory of M2, D3, M5, and D1 Branes + D5 Branes,'' hep-th/9801206.} 
\lref\skend{K. Sfetsos and K. Skenderis, ``Microscopic Derivation of
the Bekenstein-Hawking Entropy Formula for Nonextremal Black Holes,''
hep-th/9711138\semi   
H. Boonstra, B. Peeters and K. Skenderis, 
``Branes and Anti-de Sitter Space-times,'' hep-th/9801076.} 
\lref\bfss{T. Banks, W. Fischler, S. Shenker and L. Susskind,
``M Theory as a Matrix Model: A Conjecture,'' Phys. Rev. {\bf D55} (1997) 
5112.} 
\lref\dm{M. Douglas and G. Moore, ``D-branes, Quivers, and ALE
Instantons,'' hep-th/9603167.}
\lref\dgm{M. Douglas, B. Greene and D. Morrison, ``Orbifold Resolution
by D-branes,'' Nucl. Phys. {\bf B506} (1997) 84.}
\lref\ls{R. Leigh and M. Strassler, ``Exactly Marginal Operators and
Duality in Four-Dimensional ${\cal N}=1$ Supersymmetric Gauge
Theory,'' Nucl. Phys. {\bf B447} (1995) 95.}
\lref\usnext{S. Kachru and E. Silverstein, in progress.}
\lref\bz{T. Banks and A. Zaks, ``On the Phase Structure of Vectorlike
Gauge Theories with Massless Fermions,'' Nucl. Phys. {\bf B196}
(1982) 189.}
\lref\edqcd{E. Witten, ``Branes and the Dynamics of QCD,'' Nucl. Phys.
{\bf B507} (1997) 658.}
\lref\edads{E. Witten, ``Anti de Sitter Space and Holography,''
hep-th/9802150.} 
\lref\tomcc{T. Banks, ``SUSY Breaking, Cosmology, Vacuum Selection and
the Cosmological Constant in String Theory,'' hep-th/9601151.}
\lref\oferetal{O. Aharony, J. Sonnenschein, S. Yankielowicz and
S. Theisen ``Field Theory Questions for String
Theory Answers,'' Nucl. Phys. {\bf B493} (1997) 177.}
\lref\osixteen{L. Alvarez-Gaume, P. Ginsparg, G. Moore, and C.
Vafa, ``An O(16)$\times$ O(16) Heterotic String,'' Phys. Lett. 
{\bf B171} 
(1986) 155\semi
L. Dixon and J. Harvey, ``String Theories in 
Ten Dimensions Without Spacetime Supersymmetry,'' 
Nucl. Phys. {\bf B274} (1986) 93.}
\lref\ds{M. Dine and E. Silverstein, ``New M-theory Backgrounds
with Frozen Moduli,'' hep-th/9712166.}
\lref\ibanez{L. Ibanez, ``A Chiral D=4, ${\cal N}$=1 String Vacuum
with a Finite Low-Energy Effective Field Theory,'' hep-th/9802103.}

\Title{\vbox{\baselineskip12pt\hbox{hep-th/9802183}
\hbox{LBNL-41440, SLAC-PUB-7756, UCB-PTH-98/12}
}}
{\vbox{\centerline{
4d Conformal Field Theories and Strings on Orbifolds}}
}
\centerline{Shamit Kachru$^{1,3}$ and Eva Silverstein$^{2,3}$ 
}
\bigskip
\bigskip
\centerline{$^{1}$Department of Physics, University of California}
\centerline{Berkeley, CA 94720}
\smallskip
\centerline{and}
\smallskip
\centerline{Ernest Orlando Lawrence Berkeley National Laboratory}
\centerline{Mail Stop 50A-5101, Berkeley, CA 94720}
\medskip
\medskip
\centerline{$^{2}$ Stanford Linear Accelerator Center}
\centerline{Stanford University}
\centerline{Stanford, CA 94309}
\medskip
\medskip
\centerline{$^{3}$ Institute for Theoretical Physics}
\centerline{University of California}
\centerline{Santa Barbara, CA 93106}
\bigskip
\medskip
\noindent
We propose correspondences between 
4d quantum field theories with ${\cal N}=2,1,0$ (super)conformal
invariance and Type IIB string theory on various orbifolds
of $AdS_5 \times S^5$.         
We argue using the spacetime string theory, 
and check using the beta functions
(exactly for ${\cal N}=2,1$ and so far at 1-loop for the gauge
couplings in the ${\cal N}=0$ case), that
these theories have conformal fixed lines.  The latter case
potentially gives well-defined non-supersymmetric vacua of
string theory, with a mechanism for making the curvature and cosmological
constant small at nontrivial string coupling.  We suggest
a correspondence between nonsupersymmetric conformal fixed lines
and nonsupersymmetric string vacua with vanishing vacuum energy.

\Date{February 1998}

\newsec{Introduction}

Part of the recent renaissance in string theory has involved
the study of string or M-theory configurations decoupled
from gravity.  The scaling arguments which decouple
gravity do not directly depend
on supersymmetry.  
It is therefore
of interest to consider nontrivial geometrical and/or brane
configurations in backgrounds with varying amounts of spacetime
supersymmetry, to learn lessons about field theories with 
different numbers of supersymmetries and the backgrounds
in which they reside.
In particular, such techniques 
are applicable to the study of nonsupersymmetric
field theories and string backgrounds.

Several recent developments have culminated in a proposal of
``duality'' between 4d ${\cal N}=4$ supersymmetric $U(N)$ 
Yang-Mills theory (realized on $N$ D3 branes) and 
Type IIB string theory on $AdS_{5}\times S^{5}$ \refs{\gubkleb,
\malda,\skend,\gubklebp,\stanford}.
The conjecture \malda\ has been further developed in 
for example \refs{\garhir,\edads}. 
Similar conjectures relate other conformal
field theories to other string or supergravity backgrounds. 
These exciting conjectures offer both a possibility of providing
a nonperturbative definition of various string backgrounds in terms of
field theories (in a way similar to M(atrix) theory \bfss),
and a way of using supergravity/string theory to learn about conformal
field theory.

In this note, we provide evidence that various 4d field theories
with 4d ${\cal N}=0,1,2$ (super)conformal invariance
can be used to provide a nonperturbative definition of IIB string
theory in orbifold backgrounds with 0, 8, and 16 supercharges.  
Our basic strategy is to study orbifolds of type IIB on $AdS_5 \times S^5$
which preserve the AdS structure but break some of the supersymmetries. 
Then as in \malda, the $SO(4,2)$ symmetry of the AdS space should 
translate into a superconformal group on the (orbifolded) D3 brane
theory.  Although from a supergravity point of view non-freely-acting
orbifolds are singular, in string theory they are not and
we can study them reliably.  

The type IIB theory in each case has a coupling constant
$g$. Gauge theories on the other hand in general
do not have a meaningful dimensionless coupling parameter
since the coupling runs.  Therefore
in order for the correspondence to work we must find that
the beta function vanishes in the field theory.  In the
cases with supersymmetry, the spacetime coupling can 
take on any value because of the impossibility of
generating a potential for the dilaton.  So for this
reason in those cases, there
should not only be a zero of the beta function, but a
fixed line (and in general a fixed hypersurface in coupling
space whose dimension is given by the number
of deformations of the spacetime string theory which preserve
the AdS structure). Furthermore, 
even in the cases without supersymmetry,
we have an AdS factor in the spacetime theory.  The AdS symmetry
should remain to all orders and nonperturbatively
in $\alpha^\prime$ of the  
type IIB string theory, though it may be spontaneously
broken by string loops. 
Since the AdS symmetry translates into the SO(4,2)
conformal symmetry in the field theory, 
in the non-supersymmetric case we expect
the beta function to vanish at leading order in $1/N$ (planar
diagrams).

This nontrivial prediction is born out by explicit
study of the orbifolded D3 brane field theories along the lines
of \refs{\dm,\dgm}.    
In the orbifolds which preserve ${\cal N}=1,2$ supersymmetry,
we can study the exact beta functions for the couplings
in the field theory. 
This allows us to 
directly verify the existence of the predicted conformal fixed
lines, and identify marginal operators with spacetime moduli like
the IIB string coupling.   

In the ${\cal N}=0$ case, though the exact beta function
is not known, we compute the one-loop beta function for the
gauge coupling and find that it vanishes!  
In cases where there is a fixed line in the nonsupersymmetric
theory, we would obtain in this way the first example of
a nonsupersymmetric string vacuum which exists for
arbitrary values of the coupling.  For appropriate
choice of $N$, keeping $g$ nontrivial, one can then tune the curvature 
to be as small as desired.  
Then in such cases one could argue that
since the CFT should exist for all $g$, 
the spacetime background should be stable for all $g$, so
a $g$-dependent vacuum
energy should not be generated.\foot{Further computations to explore
this possibility
are underway \usnext.}  In this way, we translate the cosmological 
constant problem into a problem of finding appropriate 
nonsupersymmetric conformal 
field theories with fixed lines.

Another class of interesting non-supersymmetric models
to consider is those for which there
exists a fixed point at large $N$, as in non-supersymmetric
QCD with an appropriate number of flavors \bz.  
In general such an isolated fixed point (realized on branes
in string theory as in \malda) would translate into
an isolated, stable nonsupersymmetric vacuum of string theory.
In this type of model, which can also
be realized in branes \edqcd, one has the intriguing possibility
of fixing $g$ in terms of $N$, and therefore getting constraints
on the coupling as a function of the curvature.

In \S2 we study field theories with
${\cal N}=2$ supersymmetry and in \S3 we study theories with
${\cal N}=1$ supersymmetry.  In \S4 we present a nonsupersymmetric
orbifold example and its 1-loop beta function, and in \S5 we indulge in
a discussion of the predictions of this
duality for the vacuum energy and field theory
beta functions in the nonsupersymmetric case.

\newsec{${\cal N}=2$ Supersymmetric Theories}

Let us consider the theory on $N$ Type IIB D3 branes at the  
$\IZ_k$ orbifold singularity of an $A_{k-1}$ ALE space.  
Following \dm, we can construct
the worldvolume theory by taking $kN$ D3 branes on the covering
space and doing a $\IZ_k$ projection on both the worldvolume
fields and the Chan-Paton factors.   
 
In the picture of \malda, this translates into a $\IZ_k$ orbifold
of the $AdS_5 \times S^5$ IIB supergravity solution.  The $\IZ_k$
acts only on the $S^5$ while leaving the $AdS_5$
untouched.  The resulting spacetime 
still has an $AdS_5$ factor.  Before taking the
scaling limit \malda, the threebranes sit at a point
in the transverse $\IR^6$.  The orbifold fixes a plane
in this $\IR^6$, which intersects the $S^5$ in an $S^1$.  
So on the $S^5$ there is an $S^1$ (a great circle) fixed by
the orbifold action.  Although this means that the supergravity solution
for this background is singular, the IIB string theory is 
sensible.  
The spacetime supersymmetry is
broken from 32 to 16 supercharges by the orbifold.

The presence of the $AdS_5$ factor in spacetime means that we 
still expect the D3 brane theory to have an $SO(4,2)$ conformal
invariance $\it after$ orbifolding.   In fact, we find 
a theory with $U(N)^{k}$ gauge group and matter in the

\eqn\ntwomatt{ (N,\overline{N},1,\cdots 1) \oplus (1,N,\overline{N},\cdots 1)
\oplus \cdots \oplus (\overline{N},1, \cdots, N)} 

\noindent
Therefore, each of the $k$ $U(N)$ gauge factors effectively has 
$2N$ fundamental hypermultiplets.  The ${\cal N}=2$ gauge theory
with this matter spectrum is known to be superconformal, and in fact
to have a fixed line parametrized by the gauge coupling.\foot{In this
and later cases, we ignore the $U(1)$ factors in the $U(N)$ gauge groups
which decouple from the interacting conformal theory in the infrared.}  
There are $k$ different gauge couplings, so we in fact expect a fixed
surface of dimension $k$.  If we keep all of the gauge couplings
equal, then $g_{YM}^{2}$ simply maps to the spacetime string coupling
as in \malda. 

Since there are a total of $k$ spacetime moduli ($k-1$ blow up modes
of the $\IZ_k$ singularity, and the string coupling $g$), it is natural
to associate the other $k-1$ marginal operators along the fixed surface to
the blow up modes.  This identification of 
field theory parameters 
with spacetime moduli is significantly different from the one that
occurs in the theory of D3 brane probes at an orbifold singularity,
where the blow-up modes map to FI terms in the 
worldvolume action \dm.
However, it is motivated by the following.  The $\IZ_k$ action on
$AdS_5 \times S^5$ does not act on the AdS space.  
In perturbative Type IIB string theory there is a product
of two 2d conformal field theories representing this vacuum, one
for the AdS space and one for the $S^5$.  Since the $\IZ_k$ acts
on only the $S^5$ CFT, we expect the blow up modes will also
only affect the $S^5$ CFT.  Then, since the AdS piece is 
untouched, we still expect a 4d conformal field theory on the
D3 branes after blowing up.  The FI terms break
the conformal invariance, while the $k-1$ marginal operators
we have found of course do not.  The marginal operators therefore
correspond to the blow-up modes of the type IIB 
theory on $AdS_5 \times (S^5/\IZ_k)$.  

This distinction between
the near-horizon theory and the original bulk theory
after orbifolding will become even sharper in the 
${\cal N}=1$ case.  There the orbifold will act freely on
the $S^5$ factor, while in the full theory before the
scaling of \malda\ the orbifold has a fixed point at
the origin of the transverse $\IR^6$.  Again
we will find agreement between this geometry and the number
of marginal directions in the field theory.

\newsec{${\cal N}=1$ Supersymmetric Theories}

We can find ${\cal N}=1$ field theories by studying D3 branes
at orbifold singularities of the form $\IR^6/\Gamma$ (with $\Gamma$ a
finite abelian group) as in \dgm.  Again, we can translate the 
$\Gamma$ action to an action on $AdS_5 \times S^5$, and since it
acts orthogonally to the D3 branes, it will preserve the
AdS structure.  So in these cases we expect the worldvolume
theory on $N$ D3 branes to be an ${\cal N}=1$ superconformal field
theory. 

Here we only present the simplest example.  More general cases
work out similarly.  Consider the $\IZ_3$ orbifold

\eqn\zthree{X^{1,2,3} \rightarrow e^{{2\pi i}\over 3} X^{1,2,3}} 

\noindent
where $X^{1,2,3}$ are the three (complex) spacetime coordinates 
transverse to the $N$ parallel D3 branes.  
In the full theory, before the scaling \malda, the orbifold
\zthree\ has a fixed point at the origin ($X^\mu=0$, i.e. $r=0$ in polar
coordinates).  In polar coordinates, the original space can
be thought of as a sphere whose size varies as a function of
the radial coordinate $r$; the orbifold acts by rotations on the
sphere and the fixed point in this case occurs where the sphere shrinks to
zero at $r=0$.  After the scaling limit \malda, the space is
a product of the sphere times the AdS space; that is, the sphere
has constant volume over the AdS space and never shrinks to zero.
So the orbifold action in this case is free.  Since there are
no orbifold fixed points, we expect no blowup moduli.  So in
the field theory we should find a single marginal direction, corresponding
to the string coupling.  This is borne out in the following analysis.

We can construct the D3 brane gauge theory at the orbifold singularity by
considering $3N$ D3 branes on the covering space with suitable projections.
The result is a $U(N)^3$ gauge theory with chiral multiplets in the

\eqn\matter{3 \times \{ (N,\overline{N},1) \oplus (1,N,\overline{N}) \oplus
(\overline{N},1,N) \} } 

\noindent
Each of the $U(N)$ gauge groups therefore has the equivalent of
$N_F = 3N$ flavors of quarks.  
Let us call the three types of
charged matter fields $U^{\mu}, V^{\mu}, W^{\mu}$ with
$\mu = 1,2,3$.  

There is also a superpotential 
\eqn\superpot{W = h_{123}U^{1}V^{2}W^{3} + 
h_{312}V^{1}W^{2}U^{3} + h_{231}W^1 U^2 V^3 - 
h_{213}U^2 V^1 W^3 - h_{321}V^2 W^1 U^3 - h_{132}W^2 U^1 V^3} 
with all the couplings $h_{abc}\equiv h$ equal for the field theory
obtained from the orbifold \dgm. 

\noindent
We are predicting that this theory has a conformal fixed line.\foot{
The finiteness of the ${\cal N}$=1 $SU(4)^3$ theory with these matter
fields and couplings 
was also recently discussed by Ibanez \ibanez.} 
Following
the analysis of \ls, we can check this conjecture.  
In their notation, the scaling coefficient for a gauge coupling $g$
is

\eqn\scaling{A_g = -[ 3C_2(G) - \sum_i T(R_i) + \sum_i T(R_i)\gamma_i ]}

\noindent
where $i$ runs over charged matter fields in representation
$R_i$ and $\gamma_i$ are the anomalous
dimensions.  
Since we have $3N$ flavors of quarks for 
each of the three $U(N)$s in our theory, the first two
terms in \scaling\ cancel for each group, leaving a
contribution to each gauge coupling $\beta$-function
which is a linear combination of anomalous dimensions $\gamma_i$.

We wish to determine whether there is a family of fixed
points in this theory as a function of couplings, and to
determine its dimension.
Let us first consider directions in coupling space
which respect the $S_3$ global 
symmetry between the three complex coordinates.
In such directions, 
\eqn\gammeq{\gamma^{U^1}=\gamma^{U^2}=\gamma^{U^3}\equiv\gamma^U}
\eqn\gammaeqII{\gamma^{V^1}=\gamma^{V^2}=\gamma^{V^3}\equiv\gamma^V}
\eqn\gammaeqIII{\gamma^{W^1}=\gamma^{W^2}=\gamma^{W^3}\equiv\gamma^W}
so we have three independent $\gamma$ functions.

The overall coupling $h$ in \superpot\ is the only superpotential
coupling respecting the  $S_3$ global 
symmetry.  In addition we have three gauge couplings $g_i,i=1,\dots,3$.
The various $\beta$ functions satisfy:
\eqn\betah{\beta_h\propto \gamma^U+\gamma^V+\gamma^W}
\eqn\betagI{\beta_{g_1}\propto \gamma^U+\gamma^V}
\eqn\betagI{\beta_{g_2}\propto \gamma^V+\gamma^W}
\eqn\betagI{\beta_{g_3}\propto \gamma^U+\gamma^W}
There is one linear relation between the four $\beta$ functions,
so that setting them to zero gives us 3 conditions on the
four couplings.  This generically yields a fixed line.

So far we have found one marginal direction.  To check whether
there are any others, we can now relax the condition forcing
the couplings to preserve the global symmetry.  If we 
preserve permutations of the $\mu=1,2$ directions, we get
six independent $\gamma$ functions.  
Having relaxed the $S_3$ constraint to $\IZ_2$, there
are three independent superpotential couplings 
($h_{123}=h_{213}$, $h_{312}=h_{321}$, and $h_{231}=h_{132}$).
Combining these with the gauge couplings
$g_{1,2,3}$ gives six independent couplings.  So here one
generically does not expect a linear relation, and these 
symmetry-breaking deformations
are not marginal.  Similarly, relaxing the symmetry completely
gives nine independent $\gamma$ functions and nine
couplings (3 gauge couplings and 6 superpotential couplings).

So we indeed find a single marginal direction.  In the type IIB
string theory this direction corresponds to the string coupling.
There are other examples of finite ${\cal N}=1$ theories on D3-branes in 
IIB string theory (at orientifold fixed points) \oferetal, whose
near-horizon geometry presumably has a similar structure.






\newsec{Nonsupersymmetric Theories}

We will now turn to orbifolds preserving no supersymmetry.
One of the simplest examples satisfying level-matching is
to take a $\IZ_5$ action on two complex variables:
\eqn\nsmap{(X^1,X^2,X^3)\to (\alpha X^1, \alpha^3 X^2,X^3)}
where $\alpha=e^{{2\pi i}\over 5}$ and we are using the notation
of \S3.  This model is tachyon-free.

The projection yields a nonsupersymmetric $U(N)^5$ gauge theory
with the following matter content.  Let us denote the
five $U(N)$ factors by $U(N_i)$, with
a cyclic index $i=1,\dots,5$.
The projection on $X^3$ yields
one complex adjoint scalar in each $U(N_i)$.  From  
$X^1$ we get complex
scalars in the $(N_i,\bar N_{i+1})$ representation, and
from $X^2$ we get complex scalars in the 
$(N_i,\bar N_{i+2})$.  We get fermions in the
$(N_i, \bar N_{i\pm 2})$ and in the $(N_i, \bar N_{i\pm 1})$.
Note that this theory is not supersymmetric at the
level of the spectrum, even before considering interactions.
It has matter self-interactions inherited from
the ${\cal N}=4$ theory at tree level (whose superpotential in
${\cal N}=1$ language is $Tr[X^1,X^2]X^3$).  That is,
as in the ${\cal N}=1$ case discussed above \dgm, the terms in
the ${\cal N}=4$ superpotential involving only the fields surviving
the orbifold projection constitute 
the tree-level interactions of our matter fields.
This gives us a set of Yukawa couplings and quartic scalar couplings.

Consider the 1-loop $\beta$-function for the gauge coupling $g_i$
of $U(N_i)$.  The vector bosons contribute
$-{g_i^3\over{16\pi^2}}({{11N}\over 3})$.  The complex adjoint scalar $X^3$
contributes
$-{g_i^3\over{16\pi^2}}(-{N\over 3})$ while the fundamentals
and antifundamentals from $X^1$ and $X^2$ contribute
$-{g_i^3\over{16\pi^2}}(-{2N\over 3})$.  The fermions
contribute 
$-{g_i^3\over{16\pi^2}}(-{8N\over 3})$.
So the total $\beta_{g_i}$ vanish at one loop.  
It is important to check this for the Yukawa
couplings and quartic scalar couplings 
also, and to see if this persists to higher
orders in the loop expansion \usnext.  For now we note
that at the orbifold point, although we do not have
an ${\cal N}=4$ spectrum, we have interactions which at tree-level 
are given in terms of the gauge coupling $g$, since
the tree-level interactions are inherited from the
${\cal N}=4$ theory.  So we are on a special locus of measure
zero in the coupling space of the general non-supersymmetric
model.

Another case in which we should find concrete checks is
in orbifolding the proposals of \malda\ for describing
compactifications.  For example, we can consider 
nonsupersymmetric orbifolds acting on $T^4$ instead
of $\IR^4$ in type IIB string theory.  The proposal
in \malda\ for the unorbifolded theory is a 
$1+1$-dimensional SCFT living on wrapped D5-branes bound
to D1-branes.  The orbifold will reduce this
to a nonsupersymmetric CFT.  
If this CFT has a fixed line, then we can obtain in
this way a spacetime theory with vanishing vacuum
energy.  If there is an isolated fixed point, one
would find a stable nonsupersymmetric string vacuum.

\newsec{Discussion}

We have explained that by orbifolding transversely to D3-branes
in type IIB string theory, we should automatically produce field
theories with conformal
fixed lines in the ${\cal N}=2,1$ cases.  
In the ${\cal N}=0$ case, conformal symmetry
must persist at the level of planar diagrams.
We checked this exactly for the ${\cal N}=2,1$
cases and so far at the level of the
1-loop gauge coupling $\beta$-functions
in the  ${\cal N}=0$ case.  
There are examples in field theory where the coupling
space has marginal directions which break some
supersymmetry.  It is interesting to consider in this setup
whether such models can be related to spacetime backgrounds
in M theory.  Then transitions would occur that change the
number of spacetime supersymmetries.

The non-supersymmetric case is
of particular interest as a potential 
formulation of non-supersymmetric
string theory well-defined at non-zero but tunable
coupling.\foot{There
have been earlier studies of non-supersymmetric
string backgrounds such as the O(16) String \osixteen,
which generates a dilaton tadpole at one loop and
rolls to weak coupling.  A more
recent attempt \ds\ yields models at fixed coupling of order
one which are unfortunately difficult to study.}  
There are clearly many similar non-supersymmetric
models one can consider, in addition to checking further
properties of our examples.  
In general these models 
give us a concrete opportunity
to study the cosmological constant.  There are
many promising features of this setup.  

Let us first discuss the curvature of the $AdS\times (S^n/\Gamma)$.
If $g$ indeed turns out
to correspond to a fixed line in the D3-brane CFT, then
we have a conformal field theory and therefore a well-defined
spacetime theory for all values of $N$ and $g$.  
The curvature in string units takes the form
\eqn\curv{R\alpha'\sim {1\over{gN}}(1+a_1g+a_2g^2+\dots)~.}
Presumably
we can then tune $N$ to make the
overall spacetime curvature 
as small as desired while
retaining a nonzero string coupling $g$.

As mentioned before, another class of models
would come for example from brane configurations corresponding
to
large-N QCD with the right number of flavors.  These will have
a fixed point instead of a fixed line \bz, and will yield
a prediction for the curvature in terms of the coupling.

Let us now discuss the cosmological constant.
At one loop in the string theory, the cosmological constant
is proportional to a tadpole for the dilaton, and
in general quantum corrections produce a $g$-dependent
vacuum energy. However, 
a potential for the dilaton would suggest that the theory
does not make sense for arbitrary coupling.  
If the field theory has vanishing beta function order by
order in $g$ (in other words, a fixed line), then 
the CFT defining the background does make sense for
arbitrary coupling (at least in perturbation theory).
In particular, the quantum field theory is well defined and
therefore by the duality
there is no apparent instability in the spacetime
background. 
By this logic,
the dilaton potential
and therefore the cosmological constant
should not be generated.   
Therefore, we have translated the cosmological constant
problem into a problem of finding nonsupersymmetric conformal
fixed lines.  The latter problem 
can hopefully be studied concretely (especially in e.g. $1+1$
dimensions) \usnext. 
In the case of isolated fixed points, it would be interesting
to determine what quantity in the conformal field theory translates
into the spacetime vacuum energy.

\bigskip

\noindent\centerline{\bf Acknowledgements}

\medskip
We would like to thank O. Aharony, M. Douglas, 
D. Gross, A. Hashimoto, I. Klebanov, 
J. Maldacena, S. Shenker, and
many of the  
other participants in the ITP Duality workshop for interesting
discussions on this topic.  We would like to especially thank
T. Banks for correcting some of our comments about
the nonsupersymmetric case in an earlier version of this paper.  
This research was supported in part by 
NSF grant PHY-94-07194 through the Institute for Theoretical
Physics.
S.K. is supported by NSF grant PHY-95-14797, by DOE contract
DE-AC03-76SF00098, and by a DOE Outstanding Junior Investigator
Award.  
E.S. is supported by the DOE under
contract DE-AC03-76SF00515.  We thank the Institute for
Theoretical Physics at UCSB for
hospitality during this project.

\listrefs
\end